\documentclass[aps,prb,twocolumn,superscriptaddress]{revtex4}

\usepackage{graphicx}
\usepackage{natbib}
\usepackage{amsmath}
\usepackage{mathrsfs}

\usepackage{subfigure}
\usepackage{epsfig} 
\usepackage{epstopdf}
\usepackage{float} 
\usepackage{tikz}
\usetikzlibrary{shapes.arrows}
\usepackage[percent]{overpic}
\usepackage{comment}
\usepackage{pstricks}
\usepackage{setspace}
\usepackage{bm}% bold math

\begin{document}
%\bibliographystyle{revtex4}
%\input epsf.sty
%\preprint{}

\newcommand{\Rneq}{\rho_{\hbox{\scriptsize neq}}}
\newcommand{\Pneq}{\Psi_{\hbox{\scriptsize neq}}}
\newcommand{\Peq}{\Psi_{\hbox{\scriptsize eq}}}
\newcommand{\Ahat}{\hat{A}}
\newcommand{\Vhat}{\hat{V}}

\newcommand{\D}{\mathbf{\bar{D}}}
\newcommand{\dbar}{\mathbf{\bar{d}}}
\newcommand{\X}{\mathbf{\bar{X}}}
\newcommand{\lm}{\lambda_{\hbox{\scriptsize max}}}
\newcommand{\dm}{\dbar_{\hbox{\scriptsize max}}}
\newcommand{\Nnn}{N_{\hbox{\scriptsize nn}}}
\newcommand{\Mhat}{\hat{M}}
\newcommand{\Uhat}{\hat{U}}
\newcommand{\Xhat}{\hat{X}}
\newcommand{\Zhat}{\hat{Z}}
\newcommand{\Rhat}{\hat{R}}
\newcommand{\Hm}{{\cal H}}
\newcommand{\Tr}{\hbox{Tr}}

\title{Absence of exponential sensitivity to small perturbations in nonintegrable systems of spins 1/2}

%\title{Sensitivity of Loschmidt echoes to small perturbations in classical and quantum spin systems}

\author{B. V. Fine}
\thanks{Corresponding author}
\email{B.Fine@thphys.uni-heidelberg.de}
\address{Institute for Theoretical Physics, University of Heidelberg, Philosophenweg 19, 69120 Heidelberg, Germany}

\author{T. A. Elsayed}
\address{Institute for Theoretical Physics, University of Heidelberg, Philosophenweg 19, 69120 Heidelberg, Germany}

\author{C. M. Kropf}
\address{Institute for Theoretical Physics, University of Heidelberg, Philosophenweg 19, 69120 Heidelberg, Germany}
\address{Institute of Physics, University of Freiburg, Hermann-Herder-Str. 3, 79104 Freiburg, Germany }

\author{A. S. de Wijn}
\address{Department of Physics, Stockholm University, 106 91 Stockholm, Sweden}


\date{May 24, 2013}

\begin{abstract}

%We show that macroscopic nonintegrable lattices of spins 1/2, which are often considered to be chaotic, do not exhibit the basic property of classical chaotic systems, namely, exponential sensitivity to small perturbations. We compare chaotic lattices of classical spins and nonintegrable lattices of spins 1/2 in terms of their magnetization responses to imperfect reversal of spin dynamics known as Loschmidt echo.  In the classical case, magnetization is exponentially sensitive to small perturbations  with characteristic exponent equal to twice the value of the largest Lyapunov exponent of the system. In the case of spins 1/2, magnetization is only power-law sensitive to small perturbations. 

We show that macroscopic nonintegrable lattices of spins 1/2, which are often considered to be chaotic, do not exhibit the basic property of classical chaotic systems, namely, exponential sensitivity to small perturbations. We compare chaotic lattices of classical spins and nonintegrable lattices of spins 1/2 in terms of their magnetization responses to imperfect reversal of spin dynamics known as Loschmidt echo.  In the classical case, magnetization exhibits exponential sensitivity to small perturbations of Loschmidt echoes, which is characterized by twice the value of the largest Lyapunov exponent of the system. In the case of spins 1/2, magnetization is only power-law sensitive to small perturbations. Our findings imply that it is impossible to define Lyapunov exponents for lattices of spins 1/2 even in the macroscopic limit. At the same time, the above absence of exponential sensitivity to small perturbations is an encouraging news for the efforts to create quantum simulators. The power-law sensitivity of spin 1/2 lattices to small perturbations is predicted to be measurable in nuclear magnetic resonance experiments.

\end{abstract}

% \pacs{05.60.Gg,75.40.Mg,75.40.Gb, 76.60.-k,31.15.V-}
%\keywords{}
\maketitle

\pagebreak[4]

%\section{Introduction}

Continuous debates on the role of chaos in the behavior of many-particle systems date back to the 19th century and predate the discovery of quantum mechanics. Despite the successes of statistical physics, the notion of chaos in many-particle quantum systems is still not fully understood. A classical system is called chaotic if its phase space trajectories exhibit exponential growth of initially small deviations between them. This growth is characterized by Lyapunov exponents\cite{Gaspard-98}.  Chaos requires nonlinear dynamics.  In contrast, quantum systems are fundamentally linear and hence nonchaotic. At the same time, the notion of chaos is frequently invoked in the foundations of quantum statistical physics\cite{Rigol-08}. 

Quantum systems are often defined to be chaotic, if their classical limit is chaotic\cite{Haake-01}. This definition, however, is problematic for spin-1/2 systems\cite{Santos-12}, which do not have classical limit. At the same time, nonintegrable systems of spins 1/2 exhibit\cite{Poilblanc-93,Pals-94} the Wigner-Dyson statistics\cite{Mehta-67} of spacings between adjacent energy levels, which is known to be a generic characteristic of quantum systems that do have chaotic classical limit\cite{Bohigas-84}. It is often expected that the correspondence between spin 1/2 systems and classical chaotic systems can be established at least at the level of ensemble-averaged properties of macroscopic observables, such as the total magnetization. In this paper, however, we show that the above quantum-classical correspondence cannot be established precisely in the situation when the classical response exhibits the quantutative signature of Lyapunov instability. We arrive to the above conclusion by analysing the behavior of the total magnetization under imperfect time-reversal known as Loschmidt echo. 

The idea that chaos affects Loschmidt echo responses of macroscopic systems was first proposed in Ref.\cite{Pastawski-00} in the context of nuclear magnetic resonance (NMR)  echo experiments on a spin 1/2 system. The authors of Ref.\cite{Pastawski-00} reported that, despite their best effort, they were not able to improve the echo response beyond a certain level. They suggested that chaos inhibits one's ability to implement perfect time reversal. Similar observations without reference to chaos were also reported earlier in Ref.\cite{Skrebnev-86}. The investigations of Ref.\cite{Pastawski-00} motivated a significant body of research on Loschmidt echoes\cite{Jalabert-01,Prosen-02A,Prosen-02,Benenti-02,Eckhardt-03,Silvestrov-03,Veble-05,Gorin-06}. However, to the best of our knowledge, no quantitative connection between chaos characteristics of {\it many-spin} systems and their {\it observable} Loschmidt echo responses has yet been proposed. 

%\subsection{Formulation of the problem}

In this paper, we first demonstrate that, for macroscopic systems of classical spins, one can extract the fundamental indicator of chaos, namely, the largest Lyapunov exponent, from the behavior of the total magnetization recovered by Loschmidt echo. If real spins were classical, the above result would resolve one of the outstanding issues of statistical physics, namely, how to obtain experimental evidence of microscopic chaos in a many-particle system\cite{Gaspard-98A,Dettmann-99,Grassberger-99,Gaspard-99}.  However, we also show that nonintegrable macroscopic systems of quantum spins 1/2 do not exhibit the above signature of chaos.

We consider lattices of $N_s$ classical spins or $N_s$ quantum spins 1/2 at high temperatures governed by the nearest-neighbor (NN) Hamiltonian
\begin{equation}
\Hm_0 =\sum_{i<j}^{\hbox{\scriptsize NN}} J_x S_{ix} S_{jx}+J_y S_{iy} S_{jy}+J_z S_{iz} S_{jz}~,
\label{H}
\end{equation}
where $J_x$, $J_y$, $J_z$ are the nearest-neighbor coupling constants, and $(S_{ix}, S_{iy}, S_{iz}) \equiv {\mathbf S}_i$ either represent three projections of the classical spin on the $i$th lattice site normalized by condition ${\mathbf S}_i^2 = 1$, or denote operators of spins 1/2. Different lattice dimensions are considered --- all with periodic boundary conditions.  

We characterize Loschmidt echo response by the difference between the values of the total magnetization for perfectly and imperfectly reversed dynamics. 
The time reversal is achieved by reversing the sign of the interaction Hamiltonian as done, e.g., in NMR magic echo experiments\cite{Rhim-71,Slichter-90,Boutis-03,Sorte-11,Morgan-12}. We consider two kinds of perturbations to perfect time reversal: (i) small {\it instantaneous} rotations of spins at the moment of time reversal; and (ii) {\it continuously present} small perturbations to the Hamiltonian of the time-reversed evolution.

In the quantum case, we are primarily interested in the perturbations that are small at the level of individual spins, so that the total magnetization remains nearly the same, but, at the same time, sufficiently many spins are perturbed, so that the overlap of the perturbed and unperturbed many-spin wave functions is negligible. Although the macroscopic limit of this setting has not yet been addressed in the literature, various aspects of the results reported below were anticipated in Refs.\cite{Rice-82,Mendes-91,Gaspard-92,Partovi-92,Ford-92,Ballentine-96,Prosen-02A}.

{\it Classical spins ---}
We parameterize the phase spaces of a classical spin lattice by vector $\X \equiv \{ S_{1x}, S_{1y}, S_{1z}, S_{2x}, S_{2y}, S_{2z}, ... \}$. Here and below, boldfaced variables with bars represent vectors in many-dimensional phase-space.  We use three projections per spin even though only two of them are independent because of the constraint $\mathbf{S}_k^2 = 1$. The difference between two nearby phase-space trajectories is denoted by vector $\D \equiv \{ \delta S_{1x}, \delta S_{1y}, \delta S_{1z}, \delta S_{2x}, \delta S_{2y}, \delta S_{2z}, ... \}$. It can also be expressed as
\begin{equation}
\D(t)\equiv \X \left( t, \X_0 + \D_0 \right) - \X \left( t, \X_0 \right) ,
\label{D}
\end{equation}
where $\X \left( t, \X_0 \right)$ is a phase space trajectory as a function of time $t$ and initial position $\X_0$, and $\X \left( t, \X_0 + \D_0 \right)$ is another trajectory initially separated from the first one by infinitesimal displacement $\D_0 \equiv \D(0)$.

A system of $N_s$ classical spins is characterized by $2N_s$ Lyapunov exponents, which can be found by solving the linear stability problem  for the growth of $\D(t)$.
We denote the 
maximum positive Lyapunov exponent as $\lm$ and the corresponding Lyapunov vector as $\dm(t)$.  A many-particle system is technically defined to be chaotic, when $\lm >0$.  
In Refs.\cite{deWijn-12,deWijn-13}, we found that, for lattices of classical spins with nearest neighbor interaction, $\lm$ is an {\it intensive} quantity, i.e. its value  is size-independent for sufficiently large lattices. It can be roughly estimated as $ \lm \approx 0.25 \sqrt{\Nnn (J_x^2 + J_y^2 + J_z^2)} $, where $\Nnn$ is the number of nearest neighbors.  

%\subsection{Manifestation of the largest Lyapunov exponent}

Our treatment below is based on the same idea as the standard numerical algorithm for computing $\lm$ \cite{Benettin-80}. Namely, we consider two phase-space trajectories $\X \left( t, \X_0 \right)$ and $\X \left( t, \X_0 + \D_0 \right)$, where $\D_0$ is a very small vector pointing in a randomly selected direction. This vector has random projections on each of the Lyapunov vectors including  $\dm(0)$. After sufficiently long time, the growth of $|\D(t)|$ is entirely dominated by $\lm$, so that $\lm$ can be obtained as
\begin{equation}
\lm =  \lim_{t \to \infty; |\D(0)| \to 0} \ {1 \over t} \ln {|\D(t)| \over |\D(0)|}
\label{lambda}
\end{equation}
In order to register this exponential growth, $|\D_0|$ should be sufficiently small, so that the projections of $\D(t)$ satisfy the inequality $\delta S_{k \mu} \ll 1$ for sufficiently long time. Once the individual projections  reach values $\delta S_{k\mu} \sim 1$, the regime of Lyapunov growth terminates. Normally, the Lyapunov growth is exponential only on average, while the instantaneous growth rates fluctuate.

In an ergodic system, the asymptotic exponential growth of $|\D(t)|$ does not depend on the choice of $\X_0$ and $\D_0$\cite{exception}. This means that, when one considers an ensemble of initial conditions $\X_0$ and/or the ensemble of perturbations $\D_0$, the ensemble-average, denoted as $\langle ... \rangle$,  also exhibits asymptotic exponential growth $\langle |\D(t)| \rangle \cong e^{\lm t}$. The time required to establish this growth is, typically, of the order of $1/\lm$ (see the supplementary material\cite{supplement}). 

%\subsection{Noise}

\begin{figure} \setlength{\unitlength}{0.1cm}
\begin{picture}(88 , 58 )
{
\put(0, 0){ \epsfig{file=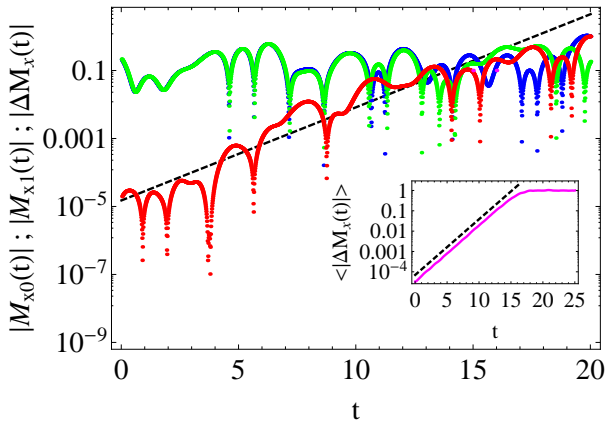,width=8.0cm } }
}
\end{picture}
\caption{  Sensitivity of classical magnetization noise to small perturbations for a cubic lattice of $16\times 16 \times 16$ classical spins with coupling constants $J_x = -0.41$, $J_y = -0.41$, $J_z = 0.82$. Blue line: magnetization noise $|M_{x0}(t)|$ for randomly chosen initial conditions $\X_0$. Green line: magnetization noise $|M_{x1}(t)|$ for the initial conditions $\X_0 + \D_0$, where $\D_0$ represents  small rotations of each spin around a random axis by an angle randomly selected from the interval $[- 10^{-4} \pi,  10^{-4} \pi] $.  Red line: $|\Delta M_x(t)| \equiv |M_{x1}(t) - M_{x0}(t) | $.  Inset: ensemble average $\langle |\Delta M_x(t)| \rangle$ over $1000$ random realizations of $\X_0$ and $\D_0$.  Black dashed lines represent $constant \times e^{\lm t}$ with $\lm = 0.63$ computed by the standard method\cite{Benettin-80,deWijn-12,supplement}.
}
\label{fig-Noise}
\end{figure}

Let us now consider the case of equilibrium noise at infinite temperature for the total $x$-component of magnetization $M_x \equiv \sum_k S_{kx}$ (see Fig.\ref{fig-Noise}). 
We compare two magnetization time series: $M_{x0}(t)$, corresponding to the initial conditions $\X_0$,  and $M_{x1}(t)$ corresponding to slightly perturbed initial conditions $\X_0 + \D_0$. In the initial small-deviations regime, $M_{x1}(t)-M_{x0}(t)$ is determined by the projection of $\D(t)$ on the direction in the phase space representing variable $M_x$ and given by the vector $\dbar_{M_x} \equiv (1,0,0,1,0,0,1, ...)$. If $\D_0$ is small enough, then there is a time interval when the growth of $\D(t)$ is controlled by $\lm$, while its orientation is controlled by $\dm(t)$. In this regime, the projection of $\D(t)$ on $\dbar_{M_x}$ fluctuates in time (and may change sign), but the amplitude of this fluctuating projection should grow exponentially as $e^{\lm t}$. As shown in Fig.\ref{fig-Noise}, this is indeed what we observed numerically\cite{exception}.  As also shown in the inset of Fig.\ref{fig-Noise}, the fluctuations can be suppressed by averaging over a large number of independent noise realizations, which means that, in the asymptotic regime,  $\langle |M_{x1}(t) - M_{x0}(t)| \rangle \cong  e^{\lm t}$.

The above analysis can now be adapted to the imperfect time reversal of magnetization noise, when one observes $M_x(t)$, and then, at time $t = t_0$, changes the sign of the Hamiltonian and simultaneously rotates each spin by a small randomly chosen angle. In this case, $M_x(t_0 - \tau)$ corresponds to  $M_{x0}(\tau)$ in the previous example and represents a perfectly time reversed signal, while $M_x(t_0 + \tau)$ corresponds to $M_{x1}(\tau)$.  Therefore, for small enough random rotations, there is a range of times $\tau$ where $\langle |M_x(t_0 - \tau) - M_x(t_0 + \tau)| \rangle \sim e^{\lm \tau}$, which, in turn, implies that
$\langle [M_x(t_0 - \tau ) - M_x(t_0 + \tau )]^2 \rangle \cong  e^{2 \lm \tau}$. 
The latter equation, together with the equilibrium relation $\langle M_x^2(t_0 - \tau )\rangle = \langle M_x^2(t_0 + \tau )\rangle \equiv \langle M_x^2 \rangle $, leads to
\begin{equation}
{ \langle  M_x(t_0 - \tau ) M_x(t_0 + \tau ) \rangle \over \langle M_x^2 \rangle } 
= 1 - C  e^{2 \lm \tau},
\label{MMe}
\end{equation}
where $C$ is a proportionality constant.

%\subsection{Relaxation}

Equation (\ref{MMe}) for equilibrium noise can now be converted into the description of a Loschmidt echo for nonequilibrium relaxation in a setting similar to NMR magic echo\cite{Slichter-90}. Namely, at $t = t_0- \tau$, the system starts in a slightly $x$-polarized state with probability distribution $\rho_0 \cong e^{- \beta M_x}$, where $\beta$ is a very small constant. For $t_0- \tau < t < t_0$, the magnetization relaxes under the action of Hamiltonian $\Hm_0$.  At $t = t_0$, the Hamiltonian switches sign, and, simultaneously, the spins are rotated by small random angles. Afterwards, the magnetization is measured at $t =t_0 + \tau$. We define the normalized echo function as 
$  F(\tau) \equiv  
\langle  M_x \rangle_{f}
/
\langle  M_x \rangle_{0}
$,
where $\langle  M_x \rangle_{0}$ and $\langle  M_x \rangle_{f}$  represent averages with respect to $\rho_0$ and $\rho_f  =   \Uhat_{-\Hm_0}(\tau ) \ \Uhat_R \ \Uhat_{\Hm_0}(\tau ) \ \rho_0$, respectively. Here, $\rho_f $ is the probability distribution at $t=t_0 + \tau$, while $\Uhat_{\Hm_0}(\tau )$ and $\Uhat_{-\Hm_0}(\tau )$ are the time evolution operator with Hamiltonians $\Hm_0$ and $-\Hm_0$, respectively, and $\Uhat_R$ is the operator representing the effect of the small spin rotations. In the limit $\beta \ll 1$, $F(\tau)$ transforms into the left-hand-side of Eq.(\ref{MMe}). Therefore, its asymptotic behavior is 
\begin{equation}
F(\tau ) = 1 - C \  e^{2 \lm \tau}.
\label{F2}
\end{equation}

\begin{figure}[t] \setlength{\unitlength}{0.1cm}
\begin{picture}(88 , 115 )
{
\put(0, 0){ \epsfig{file=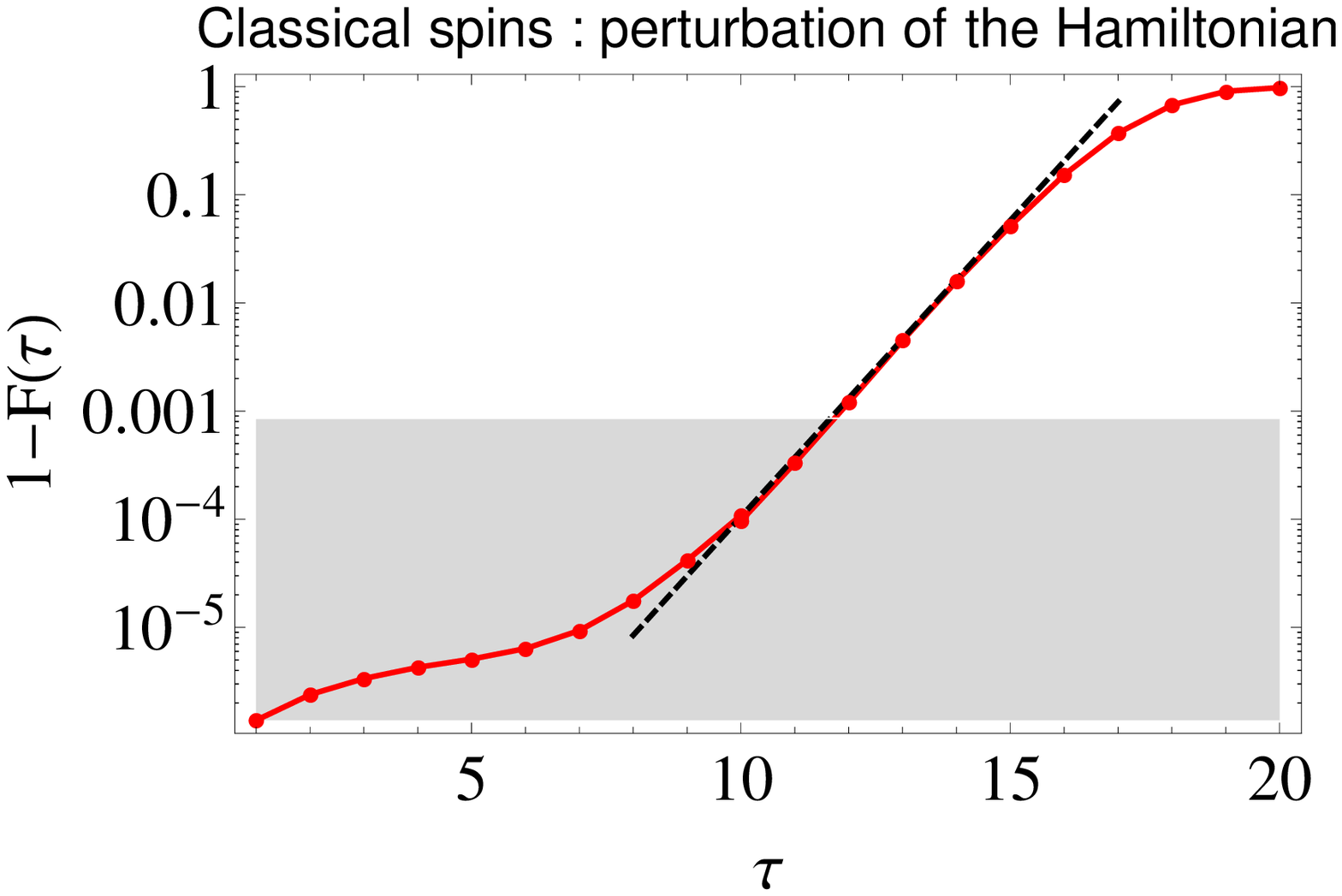,width=8cm } }
\put(0,50){(b)}
\put(0, 58){ \epsfig{file=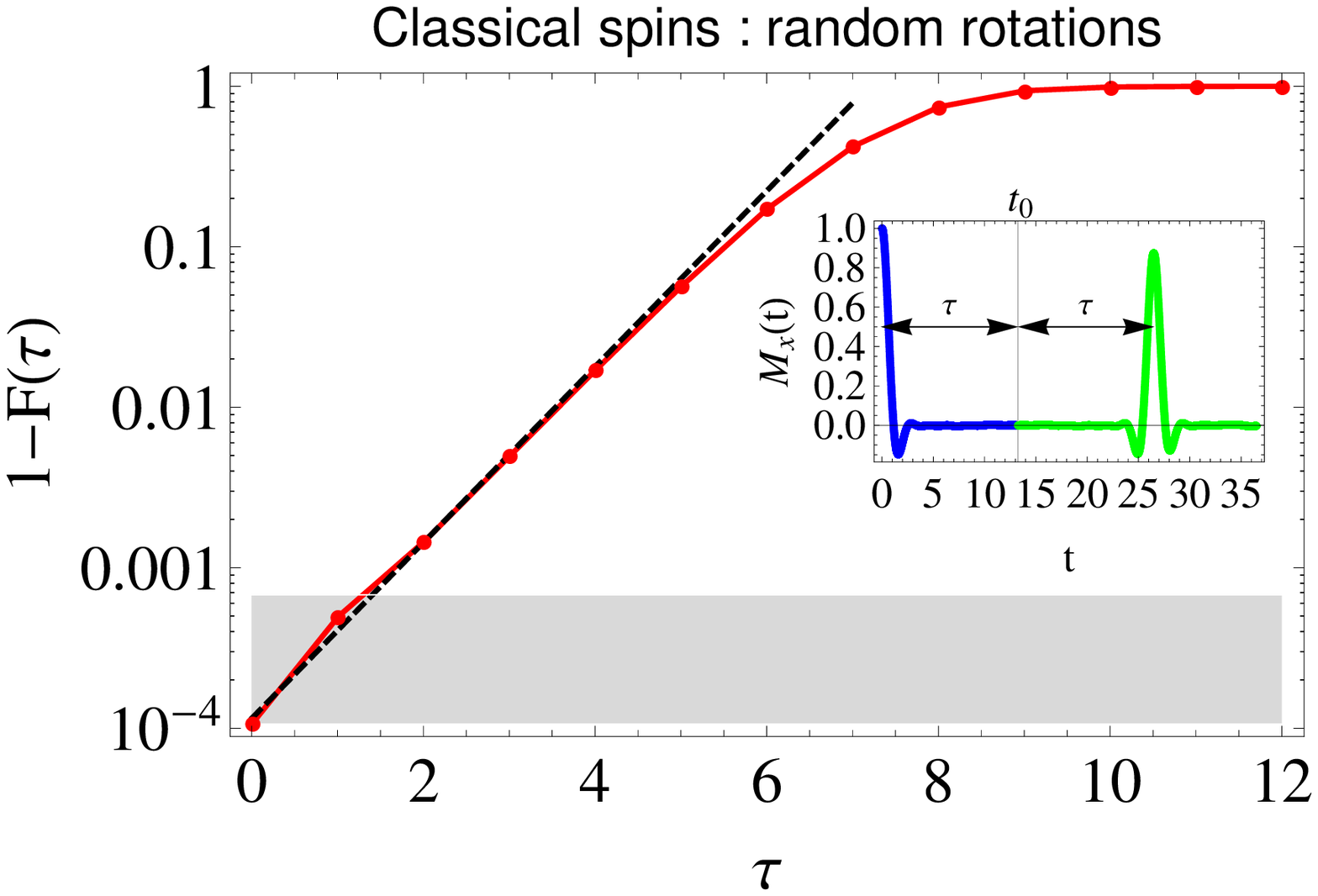,width=8cm } }
\put(0,108){(a)}
}
\end{picture}
\caption{ \label{fig-3D} Loschmidt echoes for the same lattice as in Fig.~\ref{fig-Noise}. (a) Echo disturbed at the moment of time reversal by small random rotations of spins around randomly chosen axes by angles selected from the interval $[- 10^{-2} \pi,  10^{-2} \pi] $. Inset: Relaxation and echo for one value of $\tau$. (b) Echo disturbed by the perturbation to the reversed Hamiltonian of the form $\sum_k h_k S_{kz}$, where each $h_k$ is randomly selected from the interval $[-2 \ 10^{-4}, 2 \ 10^{-4}] $.  Solid red lines represent the average over $2.8 \ 10^5$ and $1.7 \ 10^5$ independent time evolutions  in (a) and (b), respectively. The initial polarization is 10 percent.  Gray areas cover the values of $1-F(\tau)$ below four root-mean-squared values of the statistical noise for $F(\tau)$. Dashed black lines  represent $constant \times e^{2 \lm t}$ with $\lm = 0.63$.
}
\end{figure}

We have tested Eq.(\ref{F2}) numerically. The results are presented in Fig.~\ref{fig-3D}(a). They clearly exhibit the expected $e^{2 \lm \tau}$ dependence for $1 - F(\tau)$. 

Let us now consider the case when perfect time reversal is disturbed by a perturbation to the reversed Hamiltonian of the form $\sum_k h_k S_{kz}$, where $h_k$ are small random magnetic fields. Such a perturbation continuously feeds the deviation of the imperfectly reversed trajectory from the perfectly reversed one. This deviation initially grows linearly in time, but then it is exponentially amplified by the intrinsic chaotic dynamics of the unperturbed Hamiltonian as in the preceding case. Therefore, the asymptotic behavior (\ref{F2}) is also expected here.  This is, indeed, what we found numerically --- see Fig.~\ref{fig-3D}(b).

{\it Spins 1/2 ---}
Now we consider Loschmidt echo for  the relaxation of $M_x$ in spin-1/2 lattices perturbed by small random rotations around the $z$-axis at the moment of time reversal. The same linear-response relation as in the classical case allows us to express the echo function as an equilibrium correlation function\cite{supplement}
\begin{equation}
F(\tau ) =  {
\Tr \left\{ e^{i \Hm_0 \tau } \ R^{\dagger} \  e^{-i \Hm_0 \tau } \ M_x \ e^{i \Hm_0 \tau } \ R  \ e^{-i \Hm_0 \tau }  \   M_x   \right\}
\over
\Tr \left\{ M_x^2 \right\}
}.
\label{Fq}
\end{equation}
where 
\begin{equation}
R   = \prod_{k} e^{-i \delta \theta_k S_{kz} } = 
\prod_{k} [ \openone \cos(\delta \theta_k/2) - 2 i S_{kz}  \sin(\delta \theta_k/2)  ]
\label{Ra}
\end{equation}
is the quantum operator rotating each spin around the $z$-axis, and $\delta \theta_k$ are the rotation angles randomly chosen in the interval $[-\delta \theta_{max},\delta \theta_{max}]$ with $\delta \theta_{max} \ll 1$. The discussion below deals with the evolution of a typical nonequilibrium wave function representing the above trace, thereby relying\cite{supplement} on the typicality results of Refs.\cite{Bartsch-09,Elsayed-13} 

The effect of each operator $\openone \cos(\delta \theta_k/2) - 2 i S_{kz}  \sin(\delta \theta_k/2)$ in Eq.(\ref{Ra}) on a many-spin wave function is that it creates a superposition of mostly the original wave function  and a small admixture of a wave function where the $x$-projection of the $k$th spin is flipped. The probability of flipping any given spin by the action of operator $R$ is, therefore, small, but, if it happens, the value of $S_{kx}$ and hence its contribution to $M_x$ switches completely between $1/2$ and $-1/2$. The overall effect of the operator $R$ can be thought of as turning the wave function just before the time reversal, $\Psi_-$, into a superposition of wave functions $\Psi_+ = \sum_{\nu} c_{\nu} \Psi_{\nu}$, where each $\Psi_{\nu}$ is obtained from $\Psi_-$ by flipping a small fraction of randomly selected spins of the order of $<\delta \theta_k^2>$, and $c_{\nu}$ are the complex amplitudes\cite{supplement}. 

If the perfect time reversal were to be disturbed by flipping only one spin, the disturbance induced by this single spin would propagate to the neighbors as a ``perturbation bubble''. The number of perturbed spins in this bubble (quantum equivalent of $|\D(t)|$) would grow following a power law rather than an exponential. This kind of growth is not supposed to be exponential even in a chaotic classical system, because the initial perturbation is not small.   Spin-1/2 lattices accessible to direct numerical simulations are not large enough to test the above conjecture, but Loschmidt echoes disturbed by complete flipping of only one spin can be simulated for large classical spin lattices, which indeed exhibit the power-law growth of $1 - F(\tau)$\cite{supplement}. 

Let us now assume that $\Psi_+$ is equal to one of $\Psi_{\nu}$, which means that
time reversal is disturbed by flipping a small randomly selected fraction of all spins. In this case, the initial power-law disturbance around each flipped spin should grow  as the above ``perturbation bubble". Later, the bubbles around different perturbed spins would start overlapping,  and the system would enter the saturation regime $1- F(\tau) \sim 1$ without $1- F(\tau)$ ever exhibiting exponential growth.   

The fact that $\Psi_+$ is a superposition of many $\Psi_{\nu}$ does not change the above conclusion. As we show in \cite{supplement}, the interference between different $\Psi_{\nu}$ averages to zero in the expression for $F(\tau)$, which implies that echo recovery for a typical $\Psi_{\nu}$ not exhibiting the exponential growth of \mbox{$1- F(\tau)$} is representative of the entire superposition \mbox{$\Psi_+ = \sum_{\nu} c_{\nu} \Psi_{\nu}$}.

\begin{figure}[t] \setlength{\unitlength}{0.1cm}
\begin{picture}(88 , 53 )
{
%\put(0, 0){ \epsfig{file=fig_2D_OneFlip_small.eps,width=8cm } }
%\put(0,50){(b)}
\put(0, 0){ \epsfig{file=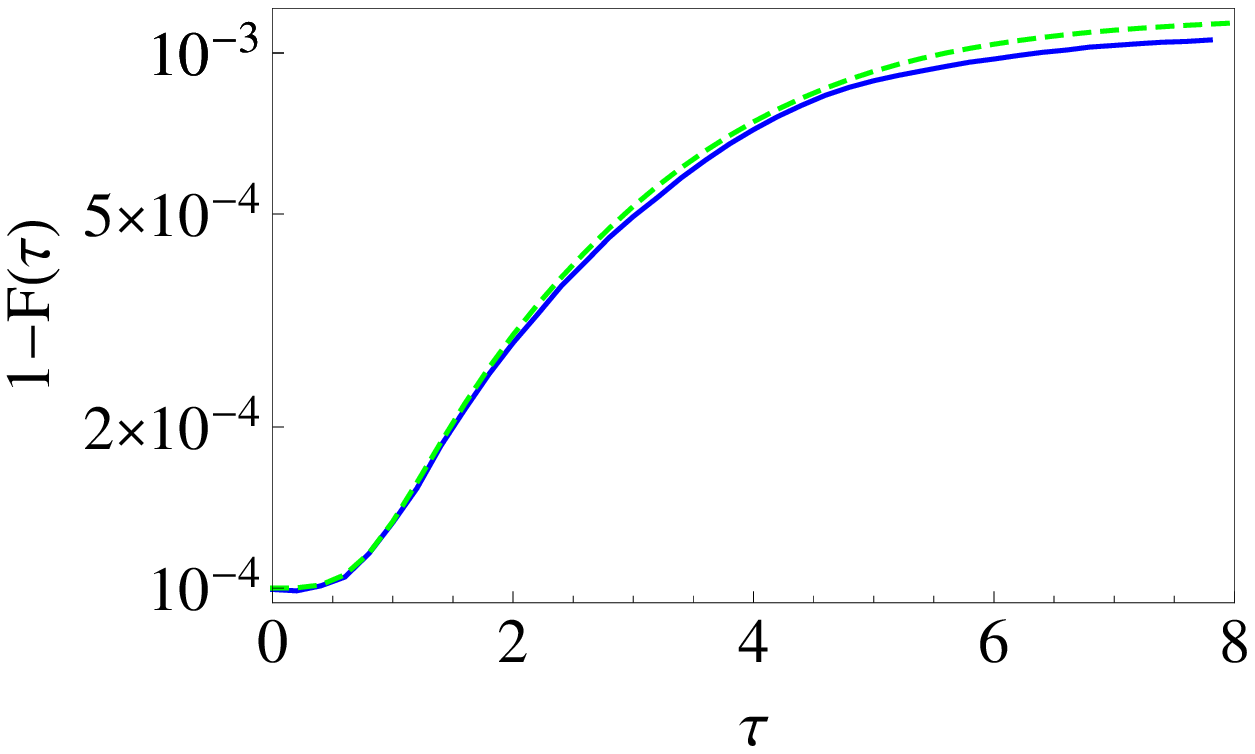,width=8.3cm } }
%\put(0,120){(a)}
%\put(20, 110){ \epsfig{file=fig_Quantum_FID.eps,width=2.2cm } }
\put(43, 12){ \epsfig{file=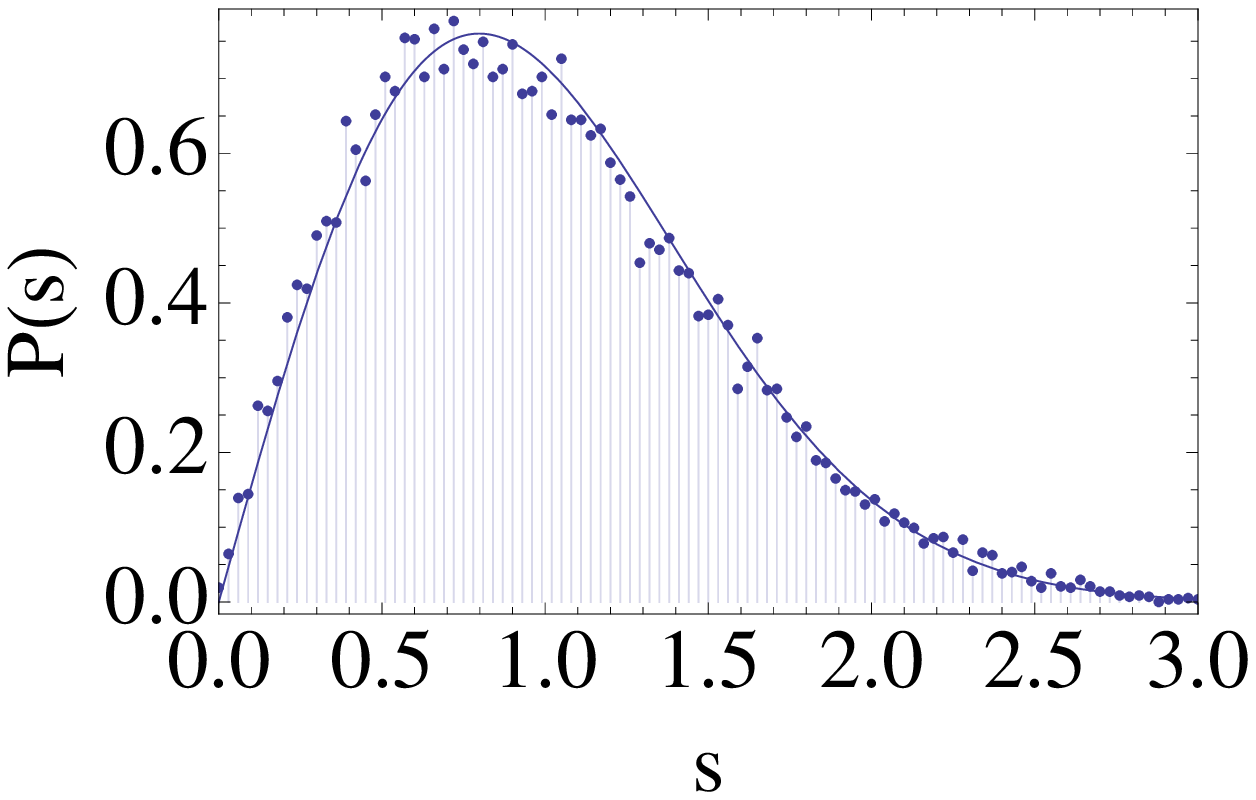,width=3.6cm } }
}
\end{picture}
\caption{  Loschmidt echoes for $5 \times 5$ square lattice of spins 1/2 with coupling constants $J_x = -0.47$, $J_y = -0.47$, $J_z = 0.94$ and 5 percent initial polarization. Blue solid line: time reversal is disturbed by small rotations around the $z$-axis with angles $\delta \theta_k$ randomly chosen from the interval $[-\pi/100, \pi/100]$. Green dashed line: time reversal is disturbed by flipping {\it one spin 1/2}, $S_{kx} \rightarrow -S_{kx}$ (plot rescaled). Inset: Evidence of nonintegrability. Dots represent the distribution $P(s)$ of level spacings $s$ for one of the irreducible blocks of $\Hm_0$. Solid line is the Wigner-Dyson fit for the Gaussian orthogonal ensemble\cite{Mehta-67}. 
}
\label{fig-Quantum}
\end{figure}

The above conclusion can, to a limited extent, be confirmed by direct quantum simulations\cite{Elsayed-13,supplement} of a nonintegrable $5 \times 5$ cluster of spins 1/2 shown in Fig.\ref{fig-Quantum}.  For this cluster, the interesting range of rotations $|\delta \theta_k| \gg 1/\sqrt{N_s}$ required to assure that $\langle \Psi_-|\Psi_+ \rangle \approx 0$  does not leave any room for a possible Lyapunov growth. Instead, we simulated the limit $|\delta \theta_k| \ll 1/\sqrt{N_s}$, which leads to $\langle \Psi_-|\Psi_+ \rangle \approx 1$ and, therefore, implies that $F(\tau)$ remains close to 1 for any $\tau$. Nevertheless, if a Lyapunov exponent were definable, $1- F(\tau)$ should have exhibited at least first signs of the exponential growth $e^{2 \lm \tau}$ before entering the saturation regime. However,  as shown in Fig.\ref{fig-Quantum}, the initial interval of quadratic growth turns immediately into subexponential growth without showing any interval of reasonably exponential behavior. 
In the same figure, we also include nearly the same Loschmidt echo shape for the case when time reversal is disturbed by flipping a single spin 1/2. In this case, the echo shape is, by definition, controlled by the growth of a single perturbation bubble.

Finally, we turn to quantum Loschmidt echo disturbed by a small perturbation to the reversed Hamiltonian. This perturbation can be viewed, by analogy with the earlier discussion for classical spins, as feeding a seed deviation between perfectly and imperfectly reversed time evolutions, which is then amplified by the intrinsic dynamics of the perfectly reversed Hamiltonian. Since, in the quantum case, this intrinsic dynamics leads to a power-law amplification, the overall echo response should exhibit a power-law sensitivity to small perturbations in the reversed Hamiltonian. The above conclusion is consistent with our finite-size simulations\cite{supplement}, but it should be properly tested in NMR magic echo experiments\cite{supplement}.

%\section{Conclusions}

%{\bf Conclusions}

To summarize, we have found that stationary nonintegrable systems of spins 1/2 do not exhibit exponential sensitivity to small perturbations of Loschmidt echoes while chaotic systems of classical spins do. This absence of exponential sensitivity in spin 1/2 systems is likely applicable beyond the Loschmidt echo setting, since it reflects the fact that extreme quantization of the projections of spins 1/2 does not leave room for the Lyapunov growth. Such a conclusion certainly represents good news for the efforts to create quantum simulators\cite{Cirac-12}, because it implies that unavoidable errors that will occur in the operation of large collections of $q$-bits will not grow exponentially in time as far as many observable quantities are concerned. At the same time, our findings are not as disturbing for the foundations of statistical physics  as they may appear at first sight. The notion of chaos defined as exponential sensitivity to small perturbations is a sufficient but not necessary condition for ergodicity, which is, in turn, required to justify Gibbs equilibrium. Also, the long-time exponential relaxation, which is known to be the same for chaotic classical and nonintegrable quantum spin systems\cite{Fine-04,Fine-03,Fine-05,Morgan-08,Sorte-11,Meier-12}, does not exclude the power-law sensitivity to small perturbations\cite{Fine-04}. 

The authors are grateful to V. Oganesyan, C. Ramanathan, H. Pastawski and V. V. Dobrovitski for related discussions. A.S.dW$'$s work is financially supported by an Unga Forskare grant from the Swedish Research Council. The numerical part of this work was performed at the bwGRiD computing cluster at the University of Heidelberg.

%\bibliography{qlyap}

\clearpage

\setcounter{figure}{0}
\renewcommand{\thefigure}{S\arabic{figure}}

\setcounter{equation}{0}
\renewcommand{\theequation}{S\arabic{equation}}

\begin{center}
{ \large \bf Supplementary material}
\end{center}

\

{\it Note:} Equation numbers, figure numbers and citation numbers appearing in this Supplementary material start with letter ``S". Equation, figure and citation numbers without ``S" refer to the text of the main article.

\

\subsection{Classical spins}
\label{classical}

\subsubsection{Numerical simulations of classical spins}

In Fig.\ref{fig-spins}, we illustrate chaotic time evolution and the onset of the Lyapunov instability for classical spin systems by an example of a ring of six interacting spins. The animation movies corresponding to Figs.\ref{fig-spins}(a) and (b) are available as online supplementary material[S1]. 

Classical spin simulations and calculations of largest Lyapunov exponents were performed as in Refs.\cite{deWijn-12,deWijn-13} using the fourth-order Runge-Kutta algorithm with discretization time step $dt \leq 0.01$. The accuracy of $\lm$ cited in the captions of Figs.~1 and 2 of the main article is $\lm = 0.63079 \pm 0.00013$. We have also performed the calculation of $\langle |\D(t)| \rangle$ for the same system. The result is shown in Fig.~\ref{fig-D}. It exhibits the features described in the main article.

\subsubsection{Loschmidt echo perturbed by a flip of one classical spin}

The main article mentions the power-law growth of $1 - F(\tau)$ for large classical spin lattices in the case when the Loschmidt echo is perturbed by a complete flip of only one spin at the moment of time reversal. Simulation results indicating these power laws are presented in Fig.~\ref{fig-onespin}. As explained in the main article, $1 - F(\tau)$ is proportional
to $\langle|\D(t)| \rangle^2$. Therefore, the power-law fits for $1 - F(\tau)$ indicated in the caption of Fig.~\ref{fig-onespin}, imply the powers 0.53, 1.15 and 1.67 for the growth of $\langle|\D(t)| \rangle$  for a chain, square lattice and cubic lattice respectively. These values are larger than, respectively, 0.5, 1.0 and 1.5 expected for a diffusive propagation of the perturbation but smaller than, respectively, 1.0, 2.0 and 3.0 expected for the ballistic propagation. The origin of the above anomalous powers is a potentially interesting subject, which, however, extends beyond the scope of the present article.

\subsection{Spins 1/2}
\label{quantum}

\begin{figure} \setlength{\unitlength}{0.1cm}
\begin{picture}(88 , 123 )
{

\put(15, 65){ \epsfig{file=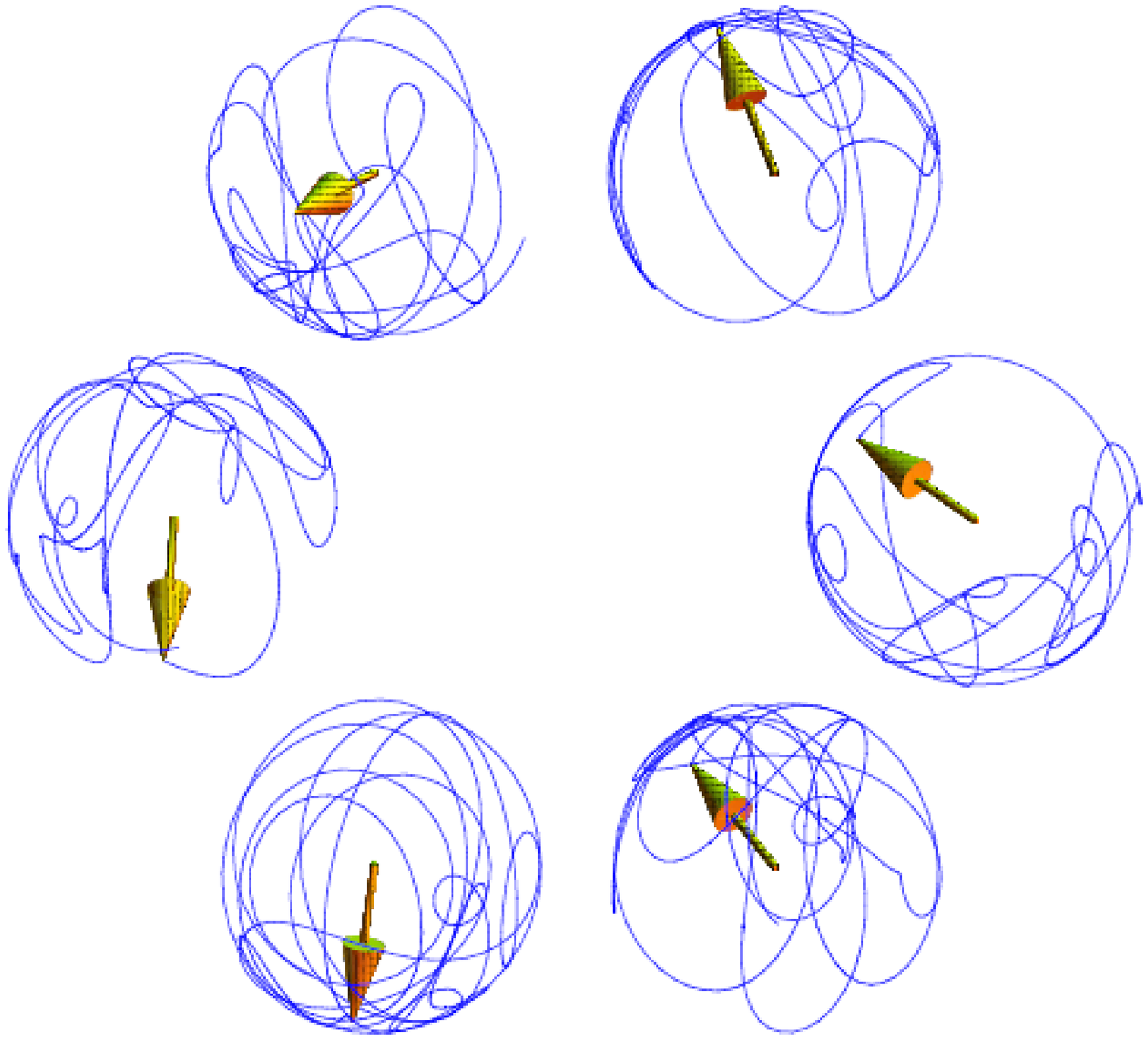,width=6cm } }
%\put(0,0){ \includegraphics[natwidth=360bp, natheight=328bp,width=7.5cm]{tmp_ClassicalSpins.png} }

\put(15, 0){ \epsfig{file=fig_Spins1.eps,width=5.4cm } }
\put(5,115){(a)}
\put(5,50){(b)}
}
\end{picture}
\caption{ \label{fig-spins}  Visualization of chaotic dynamics for a ring of six classical spins with coupling constants $J_x = -0.65$, $J_y = -0.3$, $J_z = 0.7$. (a) typical time evolution;  (b) onset of Lyapunov instability for two time evolutions with very close initial conditions. (See also the accompanying animation movies[S1].) 
}
\end{figure}

\begin{figure} \setlength{\unitlength}{0.1cm}
\begin{picture}(88 , 60 )
{
\put(0, 60){ \rotatebox{-90}{\epsfig{file=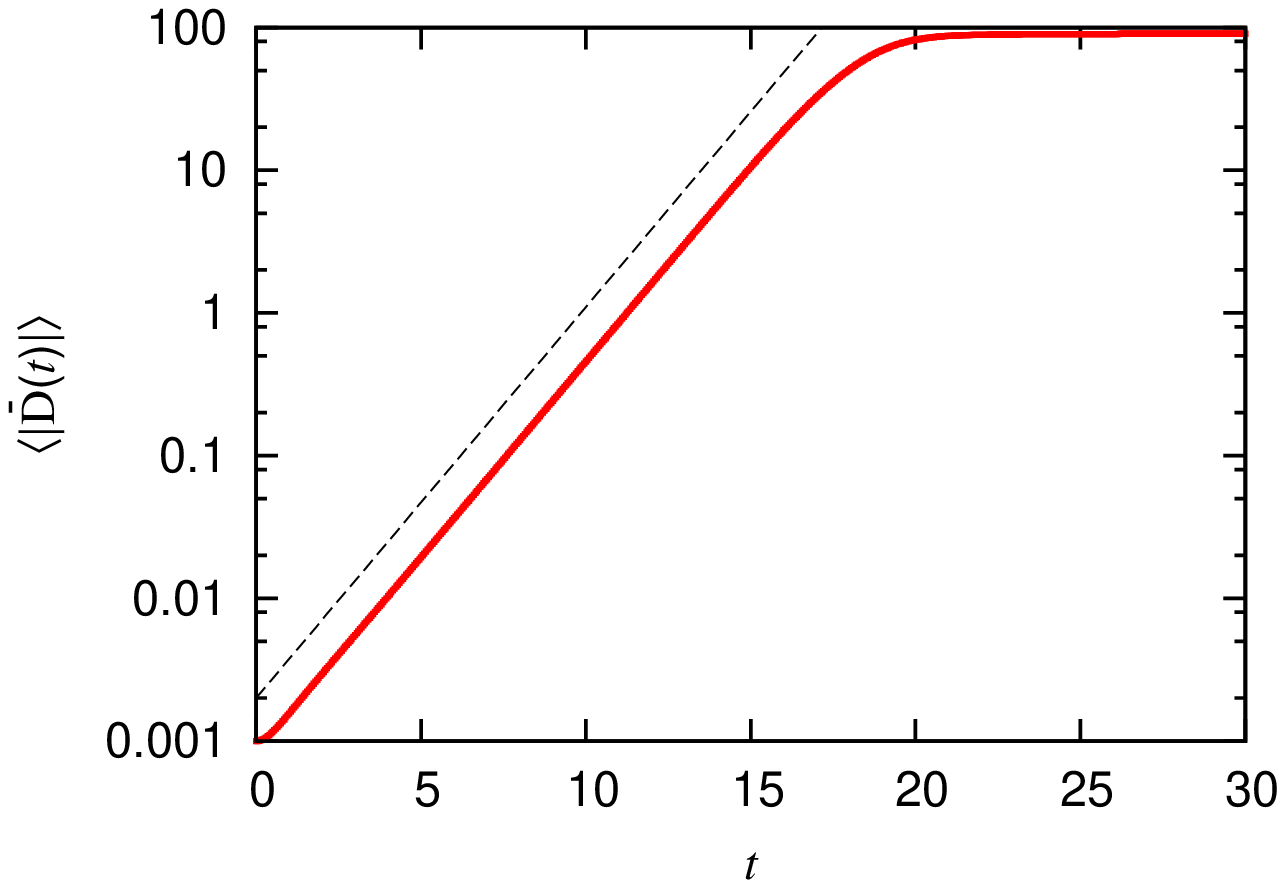,width=6cm} } }
}
\end{picture}
\caption{ \label{fig-D} Growth of $\langle|\D(t)| \rangle$ (red solid line) for a cubic lattice of $16\times 16 \times 16$ classical spins with coupling constants $J_x = -0.41$, $J_y = -0.41$, $J_z = 0.82$ averaged over 1900 pairs of phase space trajectories. The initial conditions for the first trajectory in each pair are chosen completely randomly for each spin. The initial conditions for the second trajectory are chosen by perturbing the initial conditions for the first trajectory by small random perturbations of each spin selected such that $|D(0)| = 0.001$, i.e. $\sqrt{\delta S_{k x}^2 + \delta S_{k y}^2 + \delta S_{k z}^2} \sim 10^{-5}$ . Black dashed line represents $constant \times e^{\lm t}$ with $\lm = 0.63$.
}
\end{figure} 

\begin{figure} \setlength{\unitlength}{0.1cm}
\begin{picture}(88 , 60 )
{
\put(0, 0){ \epsfig{file=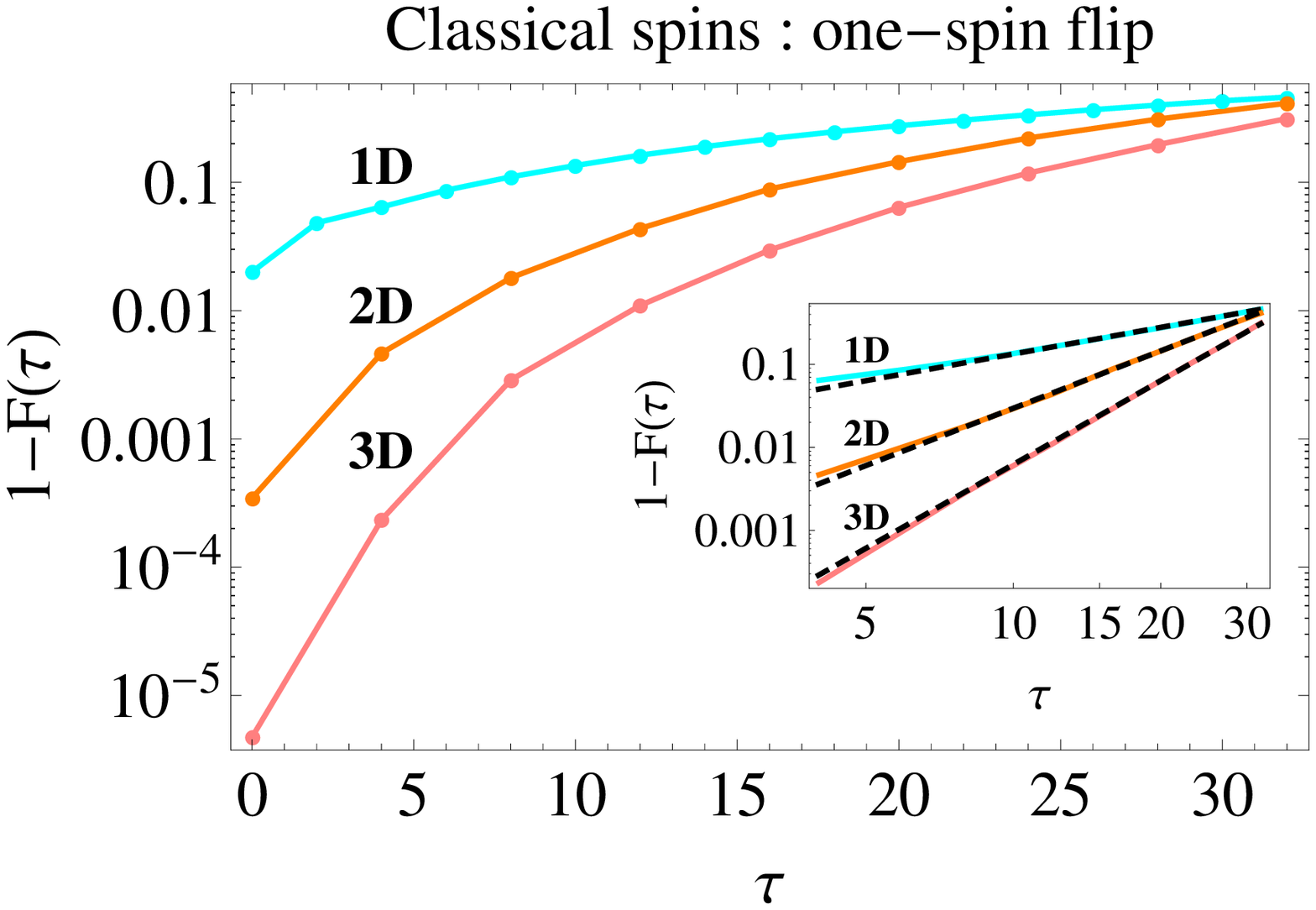,width=8cm } }
}
\end{picture}
\caption{ \label{fig-onespin} Loschmidt echoes for the relaxation of $M_x$ on lattices of classical spins disturbed at the moment of time reversal by a flip of one spin (sign change for all three spin projections). The lines represent the echoes for three different lattice dimensions: (1D) chain of 100 spins, average over $n_{e} = 4.6 \ 10^6$ time evolutions; (2D) square lattice of $100 \times 100$ spins, $n_e = 1.4 \ 10^4$, $1 - F(\tau)$ is multiplied by 2 for better visibility; (3D) cubic lattice of $64 \times 64 \times 64$ spins, $n_e = 583$.  The initial polarization in all three cases is 10 percent. Inset: log-log plot of the same results. Dashed lines: power law fits with powers 1.07, 2.30 and 3.35 for 1D, 2D and 3D, respectively.
}
\end{figure} 

\subsubsection{ Analytical considerations}

Here we elaborate on the analysis presented in the quantum part of the main article. Given the very limited sizes of spin 1/2 systems that can be simulated numerically, our claim of the absence of the exponential sensitivity of Loschmidt echoes to small perturbations is based mainly on the analytical considerations, which are, at the same time, entirely consistent with our finite-size simulations.

One aspect that distinguishes our work from many others on quantum Loschmidt echoes is that we are not concerned with quantum fidelity measuring the overlap of perfectly time-reversed and imperfectly time-reversed wave functions. Any realistically small perturbation, such as a pulse of stray magnetic fields acting on a macroscopic spin system should generate a quantum state, which, for all practical purposes, has no overlap with the original one. Instead, we are interested in the Loschmidt echo response of physical observables, such as the total magnetization.

The high-temperature fluctuation-dissipation relation allows us to express the echo function $F(\tau)$ as either an equilibrium correlation function given by Eq.(\ref{Fq}), or as a nonequilibrium relaxation function in the linear response regime\cite{Elsayed-13}:
\begin{equation}
F(\tau) \cong \Tr \left\{
 e^{i \Hm_0 \tau } \ R^{\dagger} \  e^{-i \Hm_0 \tau } \ M_x \ e^{i \Hm_0 \tau } \ R  \ e^{-i \Hm_0 \tau }  \   \rho_0   
\right\}.
\label{Fneq}
\end{equation} 
where $\rho_0 \cong e^{-\beta M_x}$ is the initial nonequilibrium density matrix.

The results on quantum typicality\cite{Bartsch-09,Elsayed-13} imply that, for systems that have a large number of quantum levels, one can substitute the exact evaluation of the trace in Eq.(\ref{Fneq}) by the evaluation of the expectation value of $M_x$ for the time evolution of a typical wave function, which, at $t=0$, is randomly selected from the ensemble of initial conditions giving on average the density matrix $\rho_0 $.   This is what we have done in our analysis and in our numerical simulations.  The accuracy of such a substitution is of the order of $2^{-N_s/2}$.

We denote the initial wave function as $\Psi_0$. The initial magnetization is then $\langle M_x \rangle_0 \equiv \langle \Psi_0 | M_x |\Psi_0 \rangle$. The wave function just before the time reversal is $\Psi_- \equiv e^{-i \Hm_0 \tau } \Psi_0$. The action of the random rotation operator converts $\Psi_-$ into 
\begin{equation}
\Psi_+ \equiv   R \Psi_- = \sum_{\nu} c_{\nu} \Psi_{\nu} ,
\label{Psi+}
\end{equation}  
as defined in the main article. The final wave function at the moment of registering the echo is  $\Psi_f = e^{i \Hm_0 \tau } \Psi_+$, while the final value of magnetization is $\langle M_x \rangle_f \equiv \langle \Psi_f | M_x |\Psi_f \rangle$. The echo function is then defined as in the classical case, namely, $F(\tau) \equiv   \langle M_x \rangle_f / \langle M_x \rangle_0 $.

In the main article, we stated that the action of operator $R$ creates a superposition of wave functions $\Psi_{\nu}$, where each of $\Psi_{\nu}$ is typically obtained from $\Psi_-$ by flipping the $x$-components of a small fraction of spins, and that no nontrivial interference between different $\Psi_{\nu}$ occurs afterwards as far as the expression for $F(\tau)$ is concerned. Let us now substantiate these statements.

Let us recall that $S_{kz} = {1 \over 2} \sigma_{kz}$, where $\sigma_{kz}$ is the Pauli matrix for the $k$th spin extended to many-spin Hilbert space. It has the property $\sigma_{kz}^2 = \openone $. We rewrite the operator $R$ as 
\begin{equation}
R= \prod_{k} [\openone \cos(\delta \theta_k/2) -  i \sigma_{kz}  \sin(\delta \theta_k/2)] = \sum_{m=0}^{N_s} A_m ,
\label{Rsigma}
\end{equation}
where $A_m$ is the sum of all possible terms  containing the product of $m$ operators $\sigma_{kz}$:
\begin{widetext}
\begin{eqnarray}
A_m &=& \left( - i \right)^m \ 
\sum_{\alpha}^{L(m)}
\left[ 
\cos (\delta \theta_{l_{\bar{\alpha} 1}}/2) 
\cos (\delta \theta_{l_{\bar{\alpha} 2}}/2)  
...
\cos (\delta \theta_{l_{\bar{\alpha}, N_s-m}}/2) 
\right]
\nonumber
\\
&& \ \ \ \ \ \ \ \ \ \ \ \ \ \ \times
 \left[ \ 
\sin (\delta \theta_{l_{\alpha 1}}/2) \sigma_{l_{\alpha 1}z} 
\sin (\delta \theta_{l_{\alpha 2}}/2) \sigma_{l_{\alpha 2}z} 
...
\sin (\delta \theta_{l_{\alpha m}}/2) \sigma_{l_{\alpha m}z} 
\right]
\nonumber
\\
& \approx &
\left( - {1 \over 2} i \right)^m \ \sum_{\alpha}^{L(m)}
\exp\left\{ 
-  {1\over 8}  \sum_p^{N_s - m} \delta \theta_{l_{\bar{\alpha} p}}^2
\right\}
\left( 
\delta \theta_{l_{\alpha 1}} \sigma_{l_{\alpha 1}z} 
\delta \theta_{l_{\alpha 2}} \sigma_{l_{\alpha 2}z} 
...
\delta \theta_{l_{\alpha m}} \sigma_{l_{\alpha m}z} 
\right)
\label{Amcos}
\end{eqnarray}
Here index $\alpha$ labels a possible combination of $m$ operators $\sigma_{kz}$: the first operator in this combination has index $k = l_{\alpha 1}$, the second $k = l_{\alpha 2}$, etc. [For an example, see Eq.(\ref{R3}) below.] Index $\bar{\alpha}$ labels the complementary set of $N_s - m$ spin operators not included in set $\alpha$. In $A_0$, the product of spin operators should be replaced by $\openone$. The approximation of the products of cosines by Gaussians in the second part of Eq.(\ref{Amcos}) is valid for large systems, when $\langle \delta \theta_k^2 \rangle \ll 1$, while $N_s \langle \delta \theta_k^2 \rangle \gg 1$.  The number of all possible combinations of $m$ spin operators is given by the binomial formula:
\begin{equation}
L(m) = {N_s! \over m! (N_s - m)!} 
\approx  {N_s! \ e^{-m(\ln m - 1)- (N_s-m)(\ln (N_s - m) - 1) } \over 2 \pi \sqrt{m (N_s - m)}} 
\label{Lm}
\end{equation}

To illustrate the expansion (\ref{Rsigma}), let us write it explicitly term-by-term for the case of three spins 1/2:
\begin{eqnarray}
R &=& \prod_{k=1}^3 [\openone \cos(\delta \theta_k/2) -  i \sigma_{kz}  \sin(\delta \theta_k/2)] =
\nonumber
\\
&&
\underbrace{\cos(\delta \theta_1/2) \cos(\delta \theta_2/2) \cos(\delta \theta_3/2) \ \openone}_{A_0} 
\nonumber
\\
&&
  - \  i \  \left[
\cos(\delta \theta_2/2) \cos(\delta \theta_3/2) \sin(\delta \theta_1/2) \ \sigma_{1z} + \cos(\delta \theta_1/2) \cos(\delta \theta_3/2)\sin(\delta \theta_2/2) \ \sigma_{2z}
\right.
\nonumber
\\
&&
 \underbrace{ \ \  \ \ \ \ \ \ \ \ \ \ \ \ \ \ \ \ \ \ \ \ \ \ \ \ \ \ \ \ \ \ \ \ \ \ \ \ \ \ \ \ \ \ \ \ \ \ \ \ \ \ \ 
\left. 
+ \  \cos(\delta \theta_1/2) \cos(\delta \theta_2/2) \sin(\delta \theta_3/2) \ \sigma_{3z}
\right]
}_{A_1}  
\nonumber
\\
&&
  + (-i)^2   \left[
\cos(\delta \theta_3/2) \sin(\delta \theta_1/2) \ \sigma_{1z} 
\ \sin(\delta \theta_2/2)  \ \sigma_{2z}  
+ 
\cos(\delta \theta_2/2) \sin(\delta \theta_1/2) \ \sigma_{1z} 
\ \sin(\delta \theta_3/2)  \ \sigma_{3z}  
\right.
\nonumber
\\
&&
 \ \  \underbrace{  \ \ \ \ \ \ \ \ \ \ \ \ \ \ \ \ \ \ \ \ \ \ \ \ \ \ \ \ \ \ \ \ \ \ \ \ \ \ \ \ \ \ \ \ \ \ \ \ \ \ \ \ \ \ 
\left. 
+ \  \cos(\delta \theta_1/2) \sin(\delta \theta_2/2) \ \sigma_{2z} 
\ \sin(\delta \theta_3/2)  \ \sigma_{3z}  
\right]
}_{A_2}  
\nonumber
\\
&&
+ \underbrace{ (-i)^3 \sin(\delta \theta_1/2) \ \sigma_{1z} \ \sin(\delta \theta_2/2) \ \sigma_{2z} 
\ \sin(\delta \theta_3/2)  \ \sigma_{3z}
}_{A_3} .
\label{R3}
\end{eqnarray}

Each wave function $\Psi_{\nu}$ introduced in Eq.(\ref{Psi+}) is obtained by the action on $\Psi_-$ by one combination of operators $\sigma_{kz}$ representing one value of $m$ and $\alpha$. In other words, index $\nu$ is a function of $m$ and $\alpha$. We can thus write 
\begin{equation}
| \Psi_{\nu(m,\alpha)} \rangle = 
\sigma_{l_{\alpha 1}z} \sigma_{l_{\alpha 2}z} ... \sigma_{l_{\alpha m}z} 
| \Psi_- \rangle
\label{Psinu}
\end{equation}
and 
\begin{equation}
c_{\nu(m,\alpha)}  \approx \left( -  {1 \over 2} i \right)^m
\exp\left\{ 
-  {1\over 8}  \sum_p^{N_s - m} \delta \theta_{l_{\bar{\alpha} p}}^2
\right\}
\delta \theta_{l_{\alpha 1}} \delta \theta_{l_{\alpha 2}} ...
\delta \theta_{l_{\alpha m}} .
\label{cnu}
\end{equation}
For $m=0$, the spins operators in Eq.(\ref{Psinu}) should be replaced by $\openone$, i.e. $| \Psi_{\nu(0,1)} \rangle = | \Psi_- \rangle $. We note here that all $\Psi_{\nu(m,\alpha)}$ are guaranteed to be normalized, because $\sigma_{kz}^2 = \openone $. 
 
Now we show that the superposition of $\sum_{\nu} c_{\nu} \Psi_{\nu}$ is dominated by wave functions $\Psi_{\nu}$ obtained by reversing the $x$-components of a small fraction of spins of the order of $\langle \delta \theta_k^2 \rangle$. Here and below, the notation $\langle ... \rangle$ denotes the average over the probability distribution of angles $\{ \delta \theta_k \}$.   We look at the average values 
\begin{equation}
\langle |c_{\nu(m,\alpha)}|^2 \rangle \approx ({1 \over 4})^m
\exp\left\{ 
-  {1\over 4}  \sum_p^{N_s - m} \langle \delta \theta_{l_{\bar{\alpha} p}}^2 \rangle
\right\}
 \langle \delta \theta_{l_{\alpha 1}}^2 \rangle \langle \delta \theta_{l_{\alpha 2}}^2 \rangle ... = e^{-(N_s - m) \kappa } \ \kappa^m,
\label{c2}
\end{equation}
where 
\begin{equation}
\kappa \equiv {1 \over 4} \langle \delta \theta_k^2 \rangle.
\label{kappa}
\end{equation}
We note that, for a given $m$, the right-hand-side of Eq.(\ref{c2}) does not depend on a particular configuration $\alpha$. 
The overall probability to find $m$ spins flipped is then given by 
\begin{equation}
P_m \approx L(m) e^{-(N_s - m) \kappa } \kappa^m 
\approx 
 {N_s! \ e^{m(\ln \kappa - \ln m + 1)- (N_s-m)[\ln (N_s - m) - 1 + \kappa] } \over 2 \pi \sqrt{m (N_s - m)}} .
\label{Pm}
\end{equation}
\end{widetext}
Neglecting the preexponential factor and maximizing the power of the exponent with respect to $m$, we find that $P_m$ has a very sharp maximum at 
\begin{equation}
m = {\kappa e^{\kappa} N_s \over \kappa e^{\kappa} + 1} \approx \kappa N_s,
\label{mmax}
\end{equation}
which means that fraction $\kappa$ of all spins is flipped in a typical $\Psi_{\nu}$.
Of course, we could have arrived to this expression simply by noting that, according to Eq.(\ref{Rsigma}), the probability for the $x$-component of each spin to be flipped is, approximately, ${1\over 4} \delta \theta_k^2$.

Now we turn to the primary quantity of interest, namely,
\begin{equation}
\langle M_x \rangle_f = \sum_{\nu,\nu'} c_{\nu'}^* c_{\nu}  \langle \Psi_{\nu'} | e^{-i \Hm_0 \tau }  \   M_x \ e^{i \Hm_0 \tau } | \Psi_{\nu} \rangle 
\label{Mf}
\end{equation}
and  show that it is determined by the diagonal terms, i.e. those with $\nu = \nu'$, 
while the off-diagonal terms 
\begin{equation}
\sum_{\nu \neq \nu'} c_{\nu'}^* c_{\nu}  \langle \Psi_{\nu'} | e^{-i \Hm_0 \tau }  \   M_x \ e^{i \Hm_0 \tau } | \Psi_{\nu} \rangle 
\label{Mfnunup}
\end{equation}
can be neglected.

For the specific case of uncorrelated random rotations around the $z$-axis, averaging over random rotation angles $\{ \delta \theta_k \}$ gives  $\langle c_{\nu'}^* c_{\nu} \rangle = 0$ in Eq.(\ref{Mfnunup}). This is because $\langle \delta \theta_k \rangle = 0$, while,  according to Eq.(\ref{cnu}), at least one factor of this kind appears in the expression for $\langle c_{\nu'}^* c_{\nu} \rangle$ with \mbox{$\nu \neq \nu'$}. The above fact implies that the average of the entire expression (\ref{Mfnunup}) is also equal to zero. In the macroscopic limit, this average should accurately represent the result for a typical single set of random angles $\{ \delta \theta_k \}$.

Our broader claim of the absence of exponential sensitivity to small perturbations does not depend on the fact that the small perturbations are caused by uncorrelated random rotations. In general, the perturbations may be correlated, which means that  $\langle c_{\nu'}^* c_{\nu} \rangle \neq 0$. However, in this case, there is a different argument: As we show below, each term $\langle \Psi_{\nu'} | e^{-i \Hm_0 \tau }  \   M_x \ e^{i \Hm_0 \tau } | \Psi_{\nu} \rangle$ in expression (\ref{Mfnunup}) cannot grow much faster than the ``perturbation bubbles" introduced in the main article to describe the growth of the diagonal terms $\langle \Psi_{\nu} | e^{-i \Hm_0 \tau }  \   M_x \ e^{i \Hm_0 \tau } | \Psi_{\nu} \rangle$. 

Let us assume that 
\begin{equation}
\Psi_{\nu} = \sigma_{q_{1}z} \sigma_{q_{2}z} ... \sigma_{p_{1}z} \sigma_{p_{2}z} ...
|\Psi_- \rangle ,
\label{Psinu1}
\end{equation}
and
\begin{equation}
\Psi_{\nu'} = \sigma_{q_{1}z} \sigma_{q_{2}z} ... \sigma_{p'_{1}z} \sigma_{p'_{2}z} ...
|\Psi_- \rangle .
\label{Psinup1}
\end{equation}
where $\{ \sigma_{q_{1}z}, \sigma_{q_{2}z}, ...\}$ is the set of operators occurring in the both expressions, while the sets $\{ \sigma_{p_{1}z}, \sigma_{p_{2}z}, ... \}$ and $\{ \sigma_{p'_{1}z}, \sigma_{p'_{2}z}, ... \}$ are only present in one of the two expressions. 

Now it is convenient to reverse the typicality argument and approximate
\begin{widetext} 
\begin{equation}
\langle \Psi_{\nu'} | e^{-i \Hm_0 \tau }  \   M_x \ e^{i \Hm_0 \tau } | \Psi_{\nu} \rangle \cong 
\Tr \left\{
e^{i \Hm_0 \tau } 
\sigma_{q_{1}z} \sigma_{q_{2}z} ... \sigma_{p'_{1}z} \sigma_{p'_{2}z} ...
e^{-i \Hm_0 \tau }  \   M_x \ e^{i \Hm_0 \tau }
\sigma_{q_{1}z} \sigma_{q_{2}z} ... \sigma_{p_{1}z} \sigma_{p_{2}z} ...
e^{-i \Hm_0 \tau }  \   M_x
\right\}
\label{trace3}
\end{equation}
\end{widetext}
The validity of the above equation can be best demonstrated by noting that, as follows from Eq.(\ref{Psinu}), there is one-to-one correspondence between wave functions $\Psi_{\nu}$ and the corresponding combinations of operators $\sigma_{kz}$ appearing in the expansion (\ref{Rsigma},\ref{Amcos}) for the operator $R$. If one substitutes this expansion for operators $R$ and $R^{\dagger}$ in Eq.(\ref{Fq}) and labels the combinations of operators $\sigma_{kz}$ by the same indices $\nu$ and $\nu'$, respectively,
then each of the resulting terms with $\nu \neq \nu'$ will correspond to $\langle \Psi_{\nu'} | e^{-i \Hm_0 \tau }  \   M_x \ e^{i \Hm_0 \tau } | \Psi_{\nu} \rangle$. (In principle, our entire analysis could have been done in terms of traces instead of typical wave functions, but such a treatment, in our opinion, would be less intuitive to present.)

The trace in the right-hand-side of Eq.(\ref{trace3}) is equal to zero when $\tau = 0$, because the set of ``unpaired'' operators $\{ \sigma_{p_{1}z} \sigma_{p_{2}z} ... \sigma_{p'_{1}z} \sigma_{p'_{2}z}... \}$ for $\nu \neq \nu'$ has at least one member. In order for the above trace to become different from zero for $\tau > 0$, the number of unpaired spins should be even, and at least pairwise correlations between them should develop under the dynamics governed by the Hamiltonian $\Hm_0$. However, since the spins belonging to each of the two original sets appearing in Eqs.(\ref{Psinu1}) and (\ref{Psinup1}) are typically distributed very sparsely on the lattice,  the unpaired subset of these spins is also typically very sparse. Therefore, the buildup of correlations between unpaired spins should be preceded by a ``dead time'' required for a perturbation of one spin to have any effect on another distant spin. Thus, for a typical pair of unpaired spins, significant correlations may emerge only by the time when the perturbation bubbles used in the main article to describe  the evolution of single $\Psi_{\nu}$ would start to overlap, and the regime $1-F(\tau) \ll 1$ is no longer valid.

\subsubsection{Numerical simulations of quantum spins.}

The Loschmidt echoes for the $5 \times 5$ square lattice of spins 1/2 shown in Fig.~\ref{fig-Quantum} as well as the quantum results presented below were computed by the method presented in Ref.\cite{Elsayed-13}. The method relies on a rigorous typicality analysis and involves the direct integration of the Schroedinger equation for a typical many-spin wave function using the fourth-order Runge-Kutta routine with the discretization time step $dt = 0.01$. The initial wave function was obtained, by, first, selecting a completely random wave function in the Hilbert space of the system and, then, subjecting that wave function to imaginary time evolution governed by operator $e^{- 0.5 \beta M_x}$.

\subsubsection{Transition to the classical limit:}

In general, for lattices of large quantum spins $S$, we expect that a Lyapunov-like growth of $1 - F(\tau)$ may only occur over the range of values of the order of $S$. This is because each spin can change its projection on a given axis only in a range between 1 and $2S$. We further note that one can think of a non-translationally invariant Hamiltonian for lattices of spins 1/2, where spins are organized into large clusters with strong coupling within a cluster and weak coupling between the clusters. In such a case, spins 1/2 within each cluster at sufficiently low temperatures can collectively mimic a large (almost classical) spin and hence a range of Lyapunov like growth may be observable for $1-F(\tau)$.

\subsubsection{Perturbed Hamiltonian}

\begin{figure}[t] \setlength{\unitlength}{0.1cm}
\begin{picture}(88 , 53 )
{

\put(0, 0){ \epsfig{file=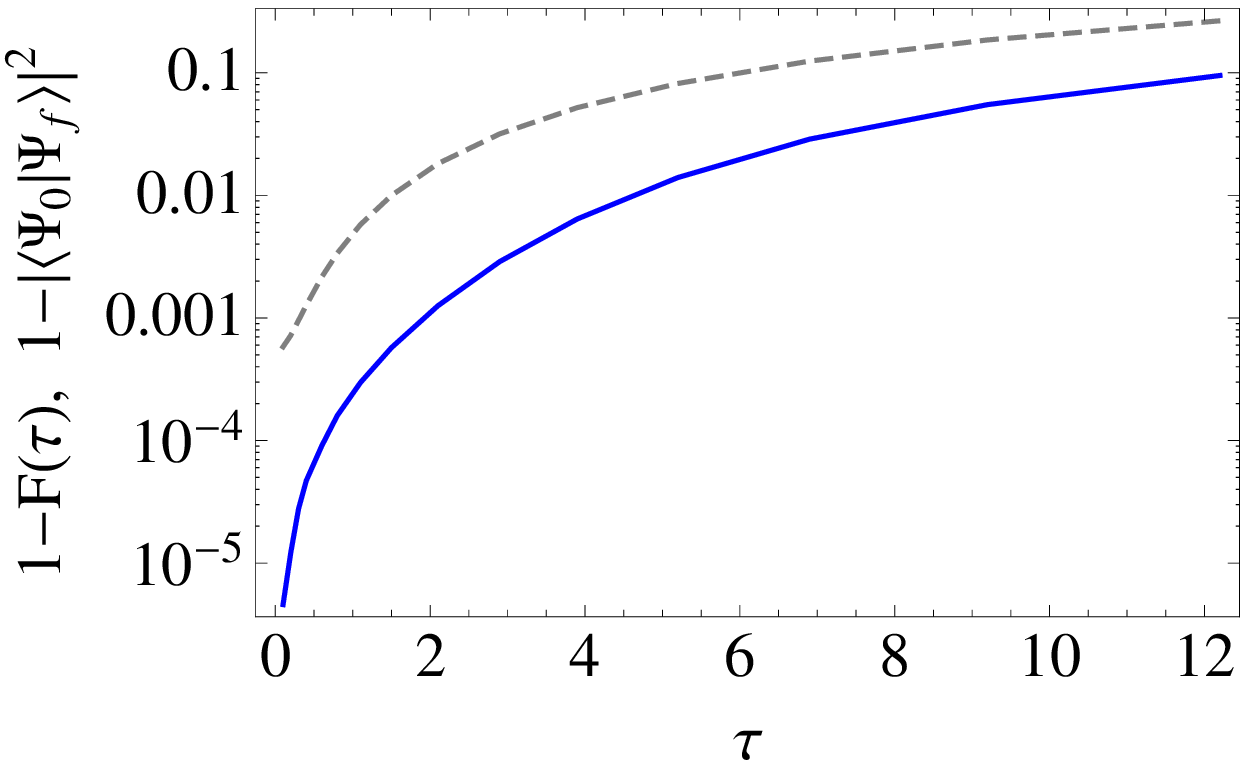,width=8.3cm } }

\put(43, 12){ \epsfig{file=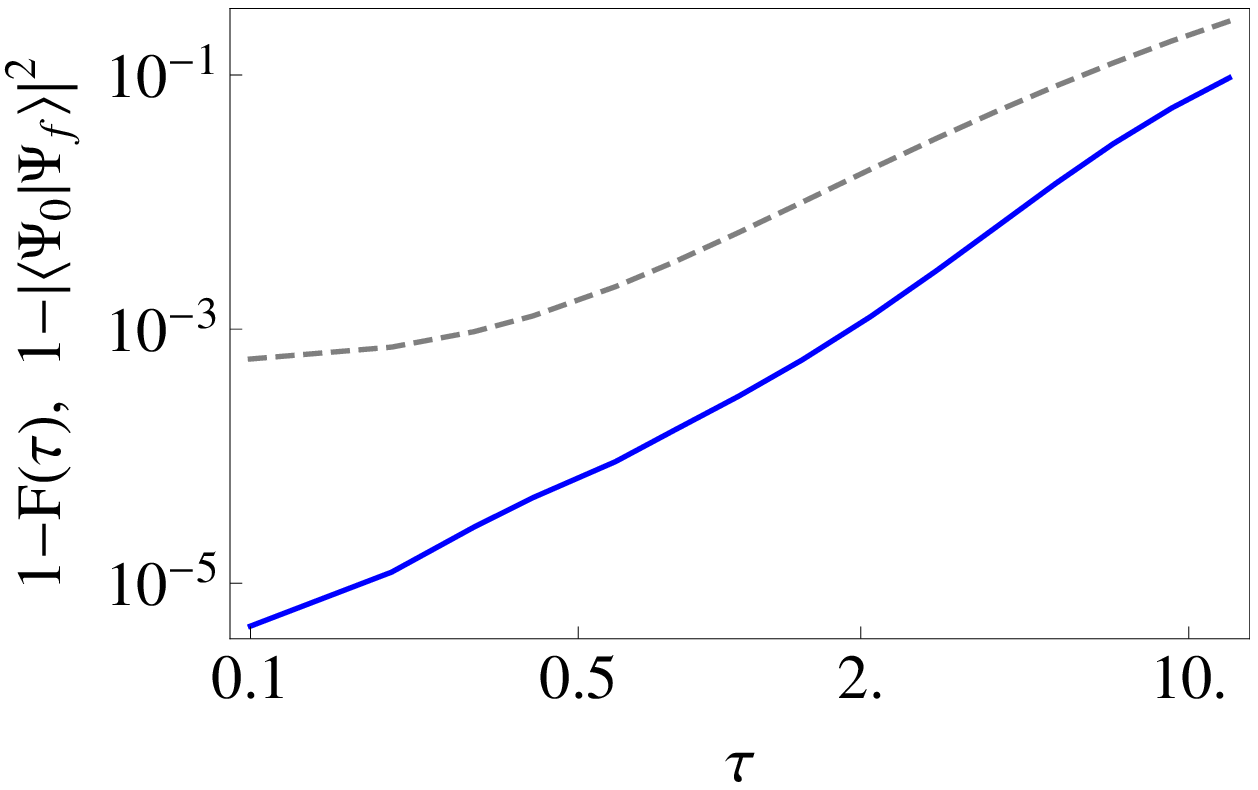,width=3.6cm } }
}
\end{picture}
\caption{ \label{fig-Quantum2} Blue line: Imperfect Loschmidt echo for the relaxation of $M_x$ on $5 \times 5$ square lattice of spins 1/2 with coupling constants $J_x = -0.47$, $J_y = -0.47$, $J_z = 0.94$ and 5 percent initial polarization. The echo is disturbed by he perturbation to the reversed Hamiltonian of the form $\sum_k h_k S_{kz}$, where each $h_k$ is randomly selected from the interval $[-0.05, 0.05] $.  Gray dashed line:   $1- |\langle \Psi_f| \Psi_0 \rangle|^2 $.
}
\end{figure}

We have also simulated the Loschmidt echo for spin 1/2 lattice perturbed by a small term in the reversed Hamiltonian. The simulations were performed for $5 \times 5$ square lattice of spins 1/2 with coupling constants $J_x = -0.47$, $J_y = -0.47$, $J_z = 0.94$ and 5 percent initial polarization. The Loschmidt echo was perturbed by adding a small term $\sum_k h_k S_{kz}$ to the reversed Hamiltonian. Each $h_k$ was randomly selected from the interval $[-0.05, 0.05] $. The result is presented in Fig.~\ref{fig-Quantum2}.  It does not exhibit any hint of an exponential growth. In this case, however, as in the case of random rotations treated in the main article, the growth of $1 - F(\tau)$ is limited by a finite size effect associated with a large overlap between  $\Psi_0$ and $\Psi_f$ --- also shown in Fig.~\ref{fig-Quantum2}. The time scale for the decay of this overlap decreases exponentially with the number of spins in the system. In the case of a macroscopic lattice, this overlap would disappear nearly instantaneously.

\subsubsection{Qualitative summary}

Here we summarize the logic of our analysis. 

We, first, have shown that Loschmidt echoes for nonintegrable lattices of spins 1/2 with short range interaction do not exhibit exponential sensitivity to small random rotations of individual spins, while the Loschmidt echoes for analogous lattices of classical spins do exhibit exponential sensitivity to analogous perturbations. Our qualitative picture is that the growth of the time reversal error caused by small random rotations of spins can be viewed as being due to an interplay of two mechanisms: local and extensive.  The local mechanism involves an initial perturbation of a given spin propagating back and forth between that spin and its neighbors. Such a perturbation is amplified every time it comes back. The extensive mechanism is associated with the growth of the overall number of spins affected directly or indirectly by the perturbation of a given one --- the so-called ``perturbation bubble." The local mechanism can lead to the exponential sensitivity to the small perturbations, but this mechanism is not available to spins 1/2, because projections of individual spins 1/2 cannot be perturbed by a small value. If a perturbation operator is applied to a given spin 1/2, spin's projection of interest is either not perturbed at all, or it is perturbed by the largest amount possible and then there is no further room for the perturbation to grow locally. (The phase coherence between the strongly perturbed and the unperturbed state quickly disappears because of the decoherence originating from the interactions with the neighbors.) On the other hand, the extensive mechanism is available to both classical and spin 1/2 lattices. However, here we rely on a conjecture that the perturbation due to the extensive mechanism cannot grow faster than a power law because of the finite maximum speed with which the perturbations can propagate. We support this conjecture by the simulations of large classical-spin lattices and finite spin-1/2 lattices.

The generalization of the above analysis to longer-range interactions and to the perturbations of the Loschmidt echoes induced by a small change in the time-reversed Hamiltonian requires one to deal with a more complicated growth of perturbations. 
Yet, in these cases, there is still no fundamental reason for the sensitivity of Loschmidt echoes to be exponential. It is removed by the fact, that projections of individual spins 1/2 cannot be perturbed by a small value.

\subsection{NMR Loschmidt echo experiments}

\subsubsection{Previous experiments}

There are several kinds of NMR echoes that fall under the broader definition of Loschmidt echoes, in particular, magic echo\cite{Rhim-71,Slichter-90,Boutis-03,Sorte-11,Morgan-12}, magic dipole echo\cite{Skrebnev-86} and polarization echo\cite{Pastawski-00}.  All of them attempt to reverse the dynamics of nuclear spins by applying strong radio-frequency (rf) fields that cause the sign change of the effective Hamiltonian of spin-spin interaction.   The basic among these echoes and also the closest to the theoretical setting of the present article is the magic echo pioneered in Ref.\cite{Rhim-71}.

As mentioned in the main article, an explicit effort to link the strength of the Loschmidt echo with the onset of microscopic chaos was made in Ref.\cite{Pastawski-00}. The authors of Ref.\cite{Pastawski-00} attempted to improve the accuracy of time reversal of polarization echo in ferrocene by increasing the strength of the rf field used in the experiment. Larger rf field implies better accuracy of the time reversal. However, Ref.\cite{Pastawski-00} and, likewise, Ref.\cite{Skrebnev-86} observed that as the strength of the rf field increases, the quality of time reversal characterized by the echo function $F(\tau)$ stops improving, which technically means that the function $F(\tau)$ stops changing. Closely related systematic investigations were also conducted very recently in Ref.~\cite{Morgan-12}.

Ref.\cite{Pastawski-00} argued that the above saturation of the echo shape implies that it is controlled by the intrinsic dynamics of the system, and therefore, since the system is expected to be chaotic, this saturation of the echo envelope is the indication of microscopic chaos. This indeed may be the case, if the Loschmidt echoes are infinitely sensitive to small perturbations. 

We, however, expect that the intrinsic dynamical sensitivity of NMR Loschmidt echoes is finite. [Here one might have a concern about the long-range character of magnetic dipolar interaction between nuclear spins, but we still expect the sensitivity of Loschmidt echoes to be finite because of the sign-changing character of the dipolar coupling constants.]  If the Loschmidt echoes exhibit finite exponential sensitivity or finite power-law sensitivity to small perturbations, then, given that the perfect time reversal leads to the perfect recovery of the echo for any time $\tau$, the shape of the Loschmidt echo is not supposed to saturate as one gradually suppresses {\it all} perturbations. 

The experimental observation of Refs.\cite{Pastawski-00,Skrebnev-86} of the echo shape independent of the strength of the rf field may be due to the experimental factors that break perfect time reversal but do not depend on the strength of the rf field. As the strength of the rf field increases and the related perturbation of the time reversal becomes smaller, eventually other perturbations take over, and hence  the echo shape becomes insensitive to the further increase of the rf field. 

\subsubsection{Proposed experiments}

While Ref.~\cite{Pastawski-00} focused on the overall shape of Loschmidt echo envelopes,
here we focus on the initial behavior of Loschmidt echoes in the regime $1 - F(\tau) \ll 1$, which has not yet been yet subjected to systematic experimental investigations.

Any experimental implementation of Loschmidt echoes, in particular, by NMR, cannot be free of small errors. In NMR, these errors are precisely of the two kinds considered in the main article: imperfect realization of NMR pulses and imperfect reversal of the effective interaction Hamiltonian.  We predict that, in NMR Loschmidt echo experiments on spin 1/2 lattices, the amplification of time reversal errors will have the character of a power law. We propose two possible ways to discriminate power-law sensitivity from  exponential sensitivity in these experiments: direct and indirect. 

The direct way is just to measure very accurately the initial behavior of $F(\tau)$ in the regime when $1-F(\tau) \ll 1$ and then see whether it exhibits a power-law or an exponential dependence over an extended time interval. Such an experiment requires very high accuracy of the measurement of the initial behavior of $F(\tau)$, which, in turn, implies very long data collection time.

The indirect way does not require the statistical accuracy of the measurements to be very high. Here one should look not at the initial behavior of $F(\tau)$ but rather at the delay with the onset of the large deviations regime $1-F(\tau) \sim 1$ as a function of a parameter $\delta$ that controls the accuracy of the time reversal. This parameter can be, for example, the inverse value of the rf field.

If $\delta$ is the only parameter controlling the quality of time reversal, then $\delta \rightarrow 0$ implies perfect time reversal. In this case, for small but finite $\delta$, we expect that, generically, $1-F(0) \sim \delta^2$. When $\delta^2$ is small enough, the initial regime of $1-F(\tau) \ll 1$ extends over sufficiently long time interval. Within this interval, $1-F(\tau)$, at first, grows quadratically as a function of $\tau$ and then transitions to either exponential or power-law asymptotic growth, dependent on the kind of sensitivity the system has. Due to the linearity of the small-perturbations regime, the time of the above transition to the asymptotic dependence should be independent of $\delta$. We denote this time as $\tau_0$. Function $1 - F(\tau)$ can now be expressed as
\begin{equation}
1 - F(\tau) = \delta^2 \times
\left\{
\begin{array}{ll}
f_0(\tau) & : \ \ \tau \leq \tau_0 \\
f_0(\tau_0) f_a(\tau- \tau_0) & : \ \ \tau > \tau_0 ,
\end{array}
\right.
\label{1mF}
\end{equation}
where $f_0(\tau)$ describes the preasymptotic stage of the growth of $1 - F(\tau)$, while $f_a(\tau)$ describes the asymptotic growth. We would like to discriminate between two asymptotic dependences
\begin{equation}
f_a(\tau- \tau_0) = e^{c_1 (\tau - \tau_1)},
\label{fa1}
\end{equation}
and 
\begin{equation}
f_a(\tau- \tau_0) = c_2 (\tau - \tau_2)^{\mu},
\label{fa2}
\end{equation}
where $c_1$, $c_2$, $\tau_1$ and $\tau_2$ are some constants.

Equation~(\ref{1mF}) is applicable as long as \mbox{$1 - F(\tau) \ll 1$}. We further expect that, once $F(\tau)$ enters the regime \mbox{$1 - F(\tau) \sim 1$}, the shape of $F(\tau)$ becomes independent of the value of $\delta$, while the $\delta$ controls only the delay time for the onset of this shape. 

We ask the question: how much time does it take for $F(\tau)$ to reach some fixed value $F_0$? (For example,  $F_0 = 0.9$.) We denote the corresponding time as $\tau_c$, and find it from equation $1- F(\tau_c) = 1 - F_0$, thereby obtaining
\begin{equation}
\tau_c = C_1 -  C_2 \ln \delta^2 
\label{tauc1}
\end{equation}
for the asymptotic dependence (\ref{fa1}), and 
\begin{equation}
\tau_c = C'_1 + C'_2 \delta^{-2/\mu} 
\label{tauc2}
\end{equation}
for the asymptotic dependence (\ref{fa2}).
Here $C_1$,$C_2$, $C'_1$ and $C'_2$  some combinations of the previously defined constants.
We expect that one can measure experimentally $\tau_c(\delta)$ and thereby discriminate the logarithmic dependence (\ref{tauc1}) from the power-law dependence(\ref{tauc2}).

If $\delta$ is not the only parameter controlling the time reversal errors, then let us assume that other errors are statistically independent of $\delta$ and characterized by parameter $\delta_0$, such that the total mean-squared error is $\delta^2 +\delta_0^2$. In this case the same treatment gives 
\begin{equation}
\tau_c = C_1 - C_2 \ln (\delta^2 + \delta_0^2) ,
\label{tauc12}
\end{equation}
for the asymptotic dependence (\ref{fa1}) and 
\begin{equation}
\tau_c = C'_1 + C'_2 (\delta^2 + \delta_0^2 )^{-1/\mu}
\label{tauc22}
\end{equation}
for the asymptotic dependence (\ref{fa2}).

Discriminating Eq.(\ref{tauc12}) from Eq.(\ref{tauc22}) involves an additional adjustable parameter $\delta_0$. This parameter can be determined from the limit of the best possible time reversal. After that, the quality of time reversal can be made systematically worse by increasing $\delta$ and thereby tracking $\tau_c(\delta)$. As long as $\delta_0^2$ is small enough, for example, of the order of $10^{-3}$, it should be possible to discriminate exponential sensitivity of Loschmidt echoes from power-law sensitivity. 

Accurate magic echo measurements in CaF$_2$ appear to be the most promising choice for the above test.

\begin{widetext}

\

\begin{samepage}

\hrulefill 

\

%\footnotesize

[S1] Links to online animation movies for Figs.~\ref{fig-spins} (a) and (b) can be found at:

\url{http://www.thphys.uni-heidelberg.de/~fine/ClassicalSpins/supplement.html} ,

\end{samepage}

\clearpage

\end{widetext}


\begin{thebibliography}{46}
\bibitem[{exc()}]{exception}
\bibinfo{note}{If $\D_0$ has no projection on $\dm(0)$, the growth of $|\D(t)|$
  is controlled by the second largest Lyapunov exponent, which is very close to
  $\lm$. In general, Lyapunov growth emerging over time interval $\Delta t$
  includes significant contributions from Lyapunov exponents $\lambda$ in the
  range $\lm - 1/\Delta t \leq \lambda \leq \lm$. This fact does not change the
  conclusions of our analysis.}
\end{thebibliography}
\end{document}